\documentclass{emulateapj}

\newcommand{\etal}{et~al.\ }

\def\omit#1{\empty}

\shorttitle{PAH Emission in Pseudobulges}

\shortauthors{D.B. Fisher}

\begin{document}
\title{Central Star Formation in Pseudobulges and Classical Bulges}

\author{
David B. Fisher \\
Department of Astronomy, University of Texas, Austin, Texas 78712;\\ dbfisher@astro.as.utexas.edu
}

\pretolerance=15000  \tolerance=15000

\begin{abstract} 
I use Spitzer 3.6-8.0 $\mu$m color profiles to compare the
radial structure of star formation in pseudobulges and classical
bulges. Pseudobulges are ``bulges'' which form through secular
evolution, rather than mergers. In this study, pseudobulges are
identified using the presence of disk-like structure in the center of
the galaxy (nuclear spirals, nuclear bars, and high ellipticity in
bulge); classical bulges are those galaxy bulges with smooth
isophotes which are round compared to the outer disk, and show no
disky structure in their bulge. I show that galaxies structurally
identified as having pseudobulges have higher central star formation
rates than those of classical bulges. Further, I also show that
galaxies identified as having classical bulges have remarkably
regular star formation profiles. The color profiles of galaxies with
classical bulges show a star forming outer disk with a sharp change,
consistent with a decline in star formation rates, toward the center
of the galaxy.  Classical bulges have a nearly constant inner profile
($r \lesssim 1.5$ kpc) that is similar to elliptical galaxies.
Pseudobulges in general show no such transition in star formation
properties from the outer disk to the central pseudobulge.  Thus I
conclude that pseudobulges and classical bulges do in fact form their
stars via different mechanisms. Further, this adds to the evidence
that classical bulges form most of their stars in fast episodic
bursts, in a similar fashion to elliptical galaxies; where as,
pseudobulges form stars from longer lasting secular processes.
\end{abstract}


\section{Introduction}

\pretolerance=15000  \tolerance=15000

Fundamental to understanding the formation of galaxies is
understanding the mechanisms responsible for forming the stars in
these galaxies. Bulges are thought to have formed their stars in and
shortly after the fast, violent process of merging stellar systems
\citep{2005sdlb.proc..143S}.  However, secular evolution can make
bulges as well.Secular evolution is the slow rearrangement of material
within a galaxy. \cite{2004ARAA..42..603K} (here after KK04) gives a
thorough review of the properties of bulges thought to be built by
secular evolution, and show many examples of pseudobulges that could
not have been made by mergers.  Secular drivers often work by causing
gas to lose angular momentum and fall to the center of the galaxy.
This effect is quite pronounced in galaxies with bars.  Hydrodynamical
simulations of gas in barred potentials, by \cite{1992MNRAS.259..345A}
show that shocks on the leading edge of bars cause this angular
momentum loss. Observations of velocity contours crowding on the
leading edges of bars supports this theory
\citep{1996ApJ...461..186D,1999ApJ...526...97R}.  It is also well
known that the surface density of star formation scales as a power law
with the surface density of gas,
$\Sigma_{SFR}\propto\Sigma^{1.4}_{gas}$ \citep{1998ApJ...498..541K}.
Therefore if enough gas is driven to the center of a disk galaxy, star
formation will convert the gas into a pseudobulge.  KK04 gives a
connection between ISM and star formation properties to a specific
kind of pseudobulge (those have star forming nuclear rings).  They
show that star formation rate densities for circumnuclear rings are
higher than their associated outer disks. Further, KK04 estimates the
timescale upon which circumnuclear disks are converted into stellar
disks, giving an estimate $\sim$0.2-2 Gyr for pseudobulge formation. 
Therefore, active star formation should be present in many present
 day pseudobulges.

Recent work on star formation in the central kiloparsec of galaxies
has shown that many bulges are forming stars, and that secular
evolution may be responsible.  \cite{2001ApJ...561..218R} compares the
radial distribution of CO to the stellar light profiles in 15 spiral
galaxies.  They find that 8 of the 15 galaxies show an excess of CO
emission in the ``bulge'' region of the galaxy, and further that the
central CO radaial distribution is similar to that of the stellar
light.  \cite{2003ApJS..145..259H} find that 45\% of the galaxies in
the BIMA SONG survey have a peak CO emission within the central
6\arcsec, while many galaxies have a central hole in the CO map.  This
suggests that there may be multiple types of molecular gas
distributions in galaxies.  Regan \etal (2001) note that this could be
the consequence of a bulge being formed via secular evolution. Stellar age 
gradients in bulges of disk galaxies suggest multiple formation mechansims as 
well. \cite{moorthy05} find that many bulges follow correlations with ellipticals. However many bulges have younger stellar populations in the center of the bulge, suggesting an outside in formation. 
\begin{figure*}[t]
\plotone{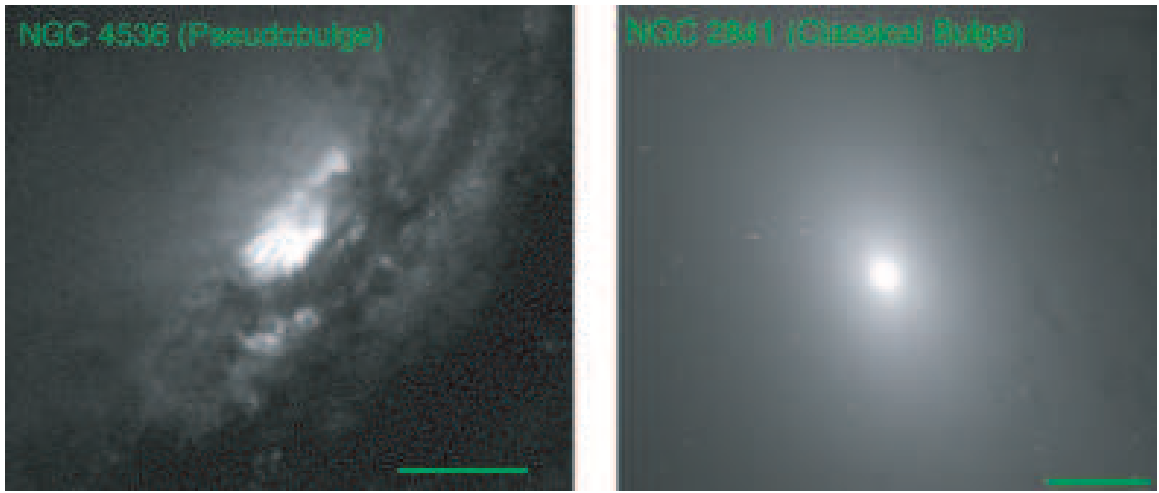}
\caption{HST images of the centers of two spiral galaxies.  On left is
an HST WFPC2 F606W image of a typical pseudobulge, NGC 4536.  On the right 
is an ACS F814W image of a galaxy
identified as a having classical bulge, NGC 2841. Notice that in the
classical bulge the over all structure is mainly featureless and
round; where as the pseudobulge is quite different it is not very
round and shows a nuclear spiral. In both images the line in the lower
right corner indicates 5 arcseconds.}
\end{figure*}

\cite{1999ApJ...525..691S} compare the concentration of 
molecular gas in the center of galaxies to the frequency of drivers
for secular evolution (i.e. bars). They find, with a sample of 20
galaxies, that molecular gas is more centrally concentrated in
galaxies with bars.  Recently, \cite{2005ApJ...632..217S} shows that
among galaxies selected to be bright in CO, barred galaxies have more
centrally concentrated gas than in galaxies without bars.  However,
they do find a few barred galaxies which do not show a large presence
of molecular gas.

In this letter I tie together indicators of star formation rates (SFR)
with other measures of secular evolution.  I show that galaxies with
any central structure indicative of pseudobulges exhibit an enhanced
amount of star formation (as indicated by the 3.6-8.0 $\mu$m color
profile and PAH emission). I also show that timescales are plausible
to assume that these pseudobulges are being built by mostly star
formation.

\section{Pseudobulge Identification} Results from HST surveys of
centers of late type galaxies \citep{2001ApJ...546..216C} have shown
that many galaxies harbor nuclear spirals, bars, and rings; these are
disk phenomena and are not possible in a hot stellar system. Also,
many spiral galaxies have similar central fattenings compared to their
outer disk \citep{1993IAUS..153..209K,2003A&A...407...61F}.  Further
Kormendy (1993) shows that many bulges have cold stellar dynamics,
more reminiscent of disks than elliptical galaxies. Many studies also
suggest that the shape of the stellar surface brightness profile of
bulges (in bulge-disk decompositions) can be used to identify
pseudobulges. Pseudobulges have nearly exponential surface brightness
profiles, as opposed to classical bulges which are closer to $r^{1/4}$
profiles \cite{1996ApJ...457L..73C, 2001ApJ...546..216C,
bashfest2005}.

KK04 ties all of this together into a single picture, suggesting
that bulges exhibiting these properties are formed through secular
evolution.  In this study, I classify galaxies as having a
pseudobulge using bulge morphology; thus if the ``bulge'' is or
contains a nuclear bar, nuclear spiral, nuclear ring, and/or the
flattening in the central region is similar to the flattening in the
outer disk, the ``bulge'' is actually a pseudobulge. Conversely if the
bulge is featureless and more round than the outer disk, the bulge is
called a classical bulge.

Figure 1 illustrates a typical example of what I identify as a pseudobulge
(left) and a classical bulge (right). Notice first that the classical
bulge (NGC 2841) has a smooth stellar light profile. There is no
reason evident in the image to think that this galaxy harbors a
pseudobulge. This galaxy has little to no dust emmission in the
bulge. Helfer \etal (2003) find no molecular gas in the center. This
classical bulge is not actively forming stars.  It is worth noting
that the presence of a little dust in the center of a galaxy does not
necessarily mean that the bulge is a pseudobulge.  
 \cite{2005AJ....129.2138L} provide many examples of nuclear
dust in elliptical galaxies, which certainly did not form through
secular evolution. Thus, merely relying on visual identification of
dust in bulges for pseudobulge identification should be done with
care.

NGC 4536 is an example of a galaxy with nuclear spiral structure and
patchiness (i.e. a pseudobulge).  A decomposition of the stellar
surface brightness profile shows that the pseudobulge dominates the
light profile out 8.5 arcseconds.  This implies that the entire
pseudobulge appears to exhibit spiral
structure. \cite{2005A&A...431..887K} find that the central 500 pc of
this galaxy is forming stars at the rate of $\sim9$ M$_{\sun}$
yr$^{-1}$ kpc$^{-2}$.

I carry out this classification process on disk galaxies in the
Spitzer archive 
data, spanning
the Hubble types S0 to Sc. I select from that only galaxies that also
have available visible band images in the the HST archive, for
pseudobulge identification. Three elliptical galaxies are added for
comparison.  Galaxies in which bulge classification is uncertain, or
those with bright AGNs are not included. The total sample is 50
galaxies.

\begin{figure*}[t]
\plotone{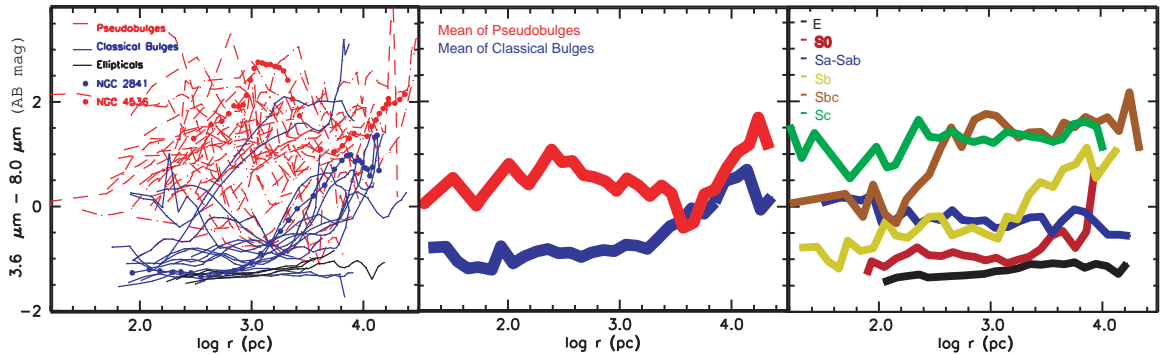}
\caption{ Profiles of all 50 galaxies considered in the sample. 
The left panel shows color profiles of all galaxies, the middle panel compares 
the averages of pseudobulges and classical bulges, and the right panel shows
the average profiles of each Hubble type.    
}
\end{figure*}

\begin{figure}[b]
\plotone{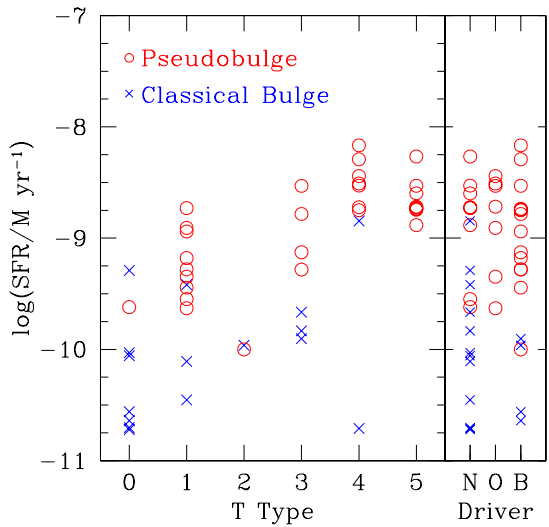}
\caption{Both panels show specific SFR of the central 1500 pc (star
formation normalized by the stellar mass).  Left panel shows
dependence of specific SFR of bulges and pseudobulges on Hubble type.
The right panel shows specific star formation of central 1500 pc
plotted against secular driving mechanism B=bar, O=oval, and N=neither
bar nor oval.  }
\end{figure}

\section{PAH emission and Color profiles}
I use the Spitzer IRAC 8 $\mu$m channel as an indicator of star
formation rates.  The usefulness of IR flux as a star formation rate
indicator has been proven by \cite{2005ApJ...632L..79W}.  They show
that luminosities a stellar light adjusted 8 $\mu$m
channel, L(PAH) = L(8$\mu$m)- 0.26 L(3.6 $\mu$m), correlate well for
giant galaxies with other star formation indicators, namely radio
luminosity and H$\alpha$ flux.

The aim of this letter is to determine if bulges that are believed to
have formed via secular evolution are more likely to be forming stars
activeley than bulges believed to have formed in mergers.  I expect to
find that galaxies which are found to harbor pseudobulges should have
a more centrally concentrated distribution of PAH emission, than
galaxies found to have classical bulges. To test this claim I
calculate surface brightness profiles of Spitzer fluxes in 3.6 $\mu$m
and 8 $\mu$m. This allows me to compare the distribution stars to that
of PAH emission in each galaxy.

The disparity in star formation properties between pseudobulges and
classical bulges is evident in the 3.6 $\mu$m - 8 $\mu$m color
profiles, shown in Figure 2 (left panel). Note that the 8$\mu$m data
in the color profile is not corrected for stellar light.  In these
color profiles regions with redder colors are more actively forming
stars (higher PAH emission per stellar luminosity).  Figure 2 shows
that those galaxies identified as having a pseudobulge do not have
markedly different color profiles in their bulges as compared to their
associated outer disks. Perhaps more compelling is the regular
behavior of the classical bulge profiles. In general, classical bulges
show a marked change in color profile getting bluer, indicating a
change toward smaller star formation rates.  The 3.6 $\mu$m - 8 $\mu$m
colors of classical bulges agree quite well with those of elliptical
galaxies (shown as black lines). Also shown (in the middle panel) is 
the averaged profile of the pseudobulges (red line) and classical bulges 
(blue line). The average profile of the pseudobulges shows a modest decline
but is roughly constant across the entire profile. Especially compared
to the average profile of the classical bulges, which shows the decrease
in star formation rates in the centers. 

Pieces of the puzzle of bulge formation are fitting together.
Those galaxies identified as pseudobulges are forming their stars 
at similar rates to outer
disks. KK04 state that pseudobulge recognition is possible because
pseudobulges have a memory of their disky past. This is generally
refering to pseudobulges having cold stellar dynamics, like disks. It
appears that this statement applies to star formation as well; the
processes which make pseudobulges are believed to be disk processes.
\cite{1997ARA&A..35..637W} remark that ``bulges are more like their
disk than they are like each other.'' Comments like this reflect a
history in which all bulge-like structures were thought to have come
from similar formation events. However, one sees clearly that
separating out pseudobulges from (merger built) classical 
bulges results in quite regular star formation properties in
classical bulges.

\section{Variation With Morphology}
 The idea that secular evolution
becomes more important at later Hubble types is well accepted (KK04
and references therein).  If pseudobulges build their mass slowly,
then it should be less likely to find a large pseudobulge because it
would take longer to form. Previous studies have found that earlier
type spirals which show molecular gas emission are emitting at higher
luminosities than later types (Sheth \etal 2005). This is actually not
surprising. Sa and S0 pseudobulges do exist (KK04). Also, the Hubble
sequence is one of decreasing bulge to disk ratio; thus secular
evolution may be either more pronounced or longer lived in galaxies
which are to become Sa galaxies. To account for variation in bulge
mass I calculate specific star formation rates (SFR normalized by the
total stellar mass of the same region). Stellar bulge masses are
caclulated by integrating the the 3.6 $\mu$m emission within 1.5 kpc,
and assuming $M/L_{3.6} \sim 1$. Figure 3 shows the specific star
formation rates for the central 1.5 kpc of each disk galaxy in my
sample. In this sample pseudobulges in Sa galaxies have roughly the
same or lower amounts of central star formation per unit mass than
later types. Though, it is worth noting that any study of Hubble types
with a sample size of 20-50 galaxies will inevitably involve small
number statistics.

The right most panel of figure 2 shows that star formation dominates
the inner kiloparsec of late type galaxies (Sbc \& Sc), has moderate 
effects of intermediate types (Sa \& Sb) and has little to no central
star formation in early types (E \& S0).
 The right panel of figure 3 shows the dependence of specific star
formation rates on secular driving mechanism. It also illustrates the
frequency of pseudobulge and bulges in galaxies with ovals (O), bars
(B), and neither (N). The result is that the average amount of central
star formation rates for galaxies with ovals and bars is about the
same.  Galaxies with a regular spiral pattern (neither bar nor oval)
 show on average lower
star formation rates. This is in agreement with the findings of
Sakamoto \etal (1999) and Sheth \etal (2005). I find, as Sheth \etal
does, that some barred galaxies exist that are not on the high end of
star formation rates. I also find that these galaxies are not
pseudobulges, possibly implying that secular evolution in galaxies
with a preexisting classical bulge is limited, or more difficult.
Another possibilty is the proposal of \cite{2005ApJ...630..837J}, that
there are mulitple stages of secular evolution. In this case early
stages are not actively forming stars. These barred non-star forming
bulges could be in the earlier (pre-starburst) stages of evolution.

\section{Conclusions} In this letter I have calculated
the PAH profiles and color profiles for 50 galaxies spanning the
Hubble types E to Sc. I use HST images to identify pseudobulges
(bulges made through secular evolution) and classical bulges (bulges
made through hierarchical mergers). I interpret the PAH emission as
being directly proportional to the star formation rate, as shown by Wu
\etal (2005). And thus I compare the incidence of active central star
formation to the presence of pseudobulge or classical bulge
structures.

Pseudobulges are shown to have higher specific star formation rates
than classical bulges. As well the PAH emission profiles of
pseudobulges are brighter and more centrally concentrated.  I also
show that galaxies with bars or ovals on average have brighter
central PAH emission than galaxies without strong drivers of secular
evolution, which is in agreement with previous findings.

As a sanity check I can calculate the time it would take the
pseudobulges in my sample to form their associated stellar
masses. This is done by simply inverting the specific star formation
rates in Figure 3.  I calclute growth times typically 0.1-5.0 Gyr,
with a median of 0.6 Gyr. Thus, assuming that the star formation is a
prolonged event, it is plausible that these galaxies have had
sufficient time to form a bulge with this amount of star
formation. And it is worth noting the similarity to the gas
consumption time scales calculated in KK04, implying that secularly
driven star formation plays a strong role in forming pseudobulges. As
well as the agreement with stellar populations work of 
\cite{2006MNRAS.366..510T} who show that many bulges ``must have experienced 
star formation events involving 10-20\% of there mass in the past 1-2 Gyr.''

The behavior of star formation rates in mergers is well studied (see
Schweizer 2005 for review); the expectation is that shocks will
induce massive star burst and exhaust available fuel relatively
quickly. Thus, the we do not expect to find merger built bulges (or
elliptical galaxies) which are actively forming stars. The response of
gas to form stars due to secular evolutoin is less well understood,
theoritically.  Secular evolution will funnel gas
inward (KK04 for review).  Therefore our finding that galaxies with
pseudobulges are much more likely to be actively forming stars is
consistent with a formation of pseudobulge via secular evolution.

\acknowledgments I wish to thank Prof. John Kormendy and Dr. Niv Drory
for helpful discussion, and prompting me to write this letter.  This
work is based in part on observations made with the Spitzer Space
Telescope, which is operated by the Jet Propulsion Laboratory,
California Institute of Technology under a contract with NASA.Some/all
of the data presented in this paper were obtained from the
Multimission Archive at the Space Telescope Science Institute
(MAST). STScI is operated by the Association of Universities for
Research in Astronomy, Inc., under NASA contract NAS5-26555. Support
for MAST for non-HST data is provided by the NASA Office of Space
Science via grant NAG5-7584 and by other grants and contracts.
\pretolerance=15000 \tolerance=15000


\end{document}